\documentclass[12pt,preprint]{aastex}
\usepackage{amsmath}
\usepackage{natbib}
\usepackage{graphicx}

\shorttitle{Lower Bounds on IGMF From Mrk 421}
\shortauthors{Takahashi et al.}

\begin{document}

\title{Lower Bounds on Magnetic Fields in Intergalactic Voids from Long-Term GeV-TeV Light Curves of the Blazar Mrk 421}

\author{Keitaro Takahashi\altaffilmark{1,7},
        Masaki Mori\altaffilmark{2},
        Kiyotomo Ichiki\altaffilmark{3},
        Susumu Inoue\altaffilmark{4,5},
        Hajime Takami\altaffilmark{6},
}

\altaffiltext{1}{Department of Physics, Kumamoto University, 2-39-1, Kurokami, Kumamoto 860-8555, Japan}
\altaffiltext{2}{Department of Physical Sciences, Ritsumeikan University, 1-1-1 Noji Higashi, Kusatsu, Shiga 525-8577, Japan}
\altaffiltext{3}{Kobayashi-Maskawa Institute for the Origin of Particles and the Universe, Nagoya University, Furo-cho, Chikusa-ku, Nagoya 464-8602, Japan}
\altaffiltext{4}{Max-Planck-Institut f\"ur Kernphysik, Saupfercheckweg 1, 69117 Heidelberg, Germany}
\altaffiltext{5}{Institute for Cosmic Ray Research, University of Tokyo, 5-1-5, Kashiwanoha, Kashiwa, Chiba 277-8582, Japan}
\altaffiltext{6}{Theory Center, Institute of Particle and Nuclear Studies, KEK, 1-1, Oho, Tsukuba 305-0801, Japan}
\altaffiltext{7}{keitaro@sci.kumamoto-u.ac.jp}

\begin{abstract}
Lower bounds are derived on the amplitude $B$ of intergalactic magnetic fields (IGMFs) in the region between Galaxy and the blazar Mrk 421, from constraints on the delayed GeV pair-echo flux that are emitted by secondary $e^-e^+$ produced in $\gamma\gamma$ interactions between primary TeV gamma-rays and the cosmic infrared background. The distribution of galaxies mapped by the Sloan Digital Sky Survey shows that this region is dominated by a large intergalactic void. We utilize data from long-term, simultaneous GeV-TeV observations by the {\it Fermi} Large Area Telescope and the ARGO-YBJ experiment extending over 850 days. For an assumed value of $B$, we evaluate the daily GeV pair-echo flux expected from the TeV data, select the dates where this exceeds the {\it Fermi} 2-$\sigma$ sensitivity, compute the probability that this flux is excluded by the {\it Fermi} data for each date, and then combine the probabilities using the inverse normal method. Consequently, we exclude $B < 10^{-20.5}~{\rm G}$ for a field coherence length of 1 kpc at $\sim$ 4-$\sigma$ level, as long as plasma instabilities are unimportant for cooling of the pair beam. This is much more significant than the 2-$\sigma$ bounds we obtained previously from observations of Mrk 501, by virtue of more extensive data from the ARGO-YBJ, as well as improved statistical analysis. Compared with most other studies of IGMF bounds, the evidence we present here for a non-zero IGMF is more robust as it does not rely on unproven assumptions on the primary TeV emission during unobserved periods.
\end{abstract}


\section{Introduction}

Intergalactic magnetic fields (IGMF), particularly those inside intergalactic void regions, have attracted much interest
as possible remnants of primordial magnetic fields that were generated in the early Universe \citep[e.g.][]{Gnedin,Langer,Takahashi1,Ichiki2}. While such fields can be amplified later within galaxies and galaxy clusters by dynamo processes, they may remain unaffected by subsequent astrophysical effects deep inside voids. Thus, IGMFs are expected to be a window onto the early Universe. For comprehensive reviews on primordial and intergalactic magnetic fields, see Widrow (2002), Widrow et al. (2012) and Ryu et al. (2012).

However, the predicted amplitudes for IGMFs of primordial origin are generally very small, $B = 10^{-25} - 10^{-15}~{\rm G}$, and difficult to probe through Faraday rotation measurements in distant radio sources or their effects on the anisotropy of the cosmic microwave background (CMB). In this context, a method that is sensitive to weak IGMFs utilizing delayed secondary emission from high-energy gamma-ray sources was proposed by Plaga (1995) and subsequently developed by many authors \citep{Dai,Razzaque,Murase1,Murase2,Ichiki1,Takahashi4,Elyiv,Neronov1,Takahashi2}. Such emission that we refer to as ``pair echos'' is expected to occur typically at GeV energies, for which the {\it Fermi} Large Area Telescope (LAT) is currently the most sensitive instrument. Since the echo flux is predicted to be larger for smaller $B$, a GeV upper limit on such components translates into a lower bound on $B$.

\begin{figure}[t]
\epsscale{1}
\plotone{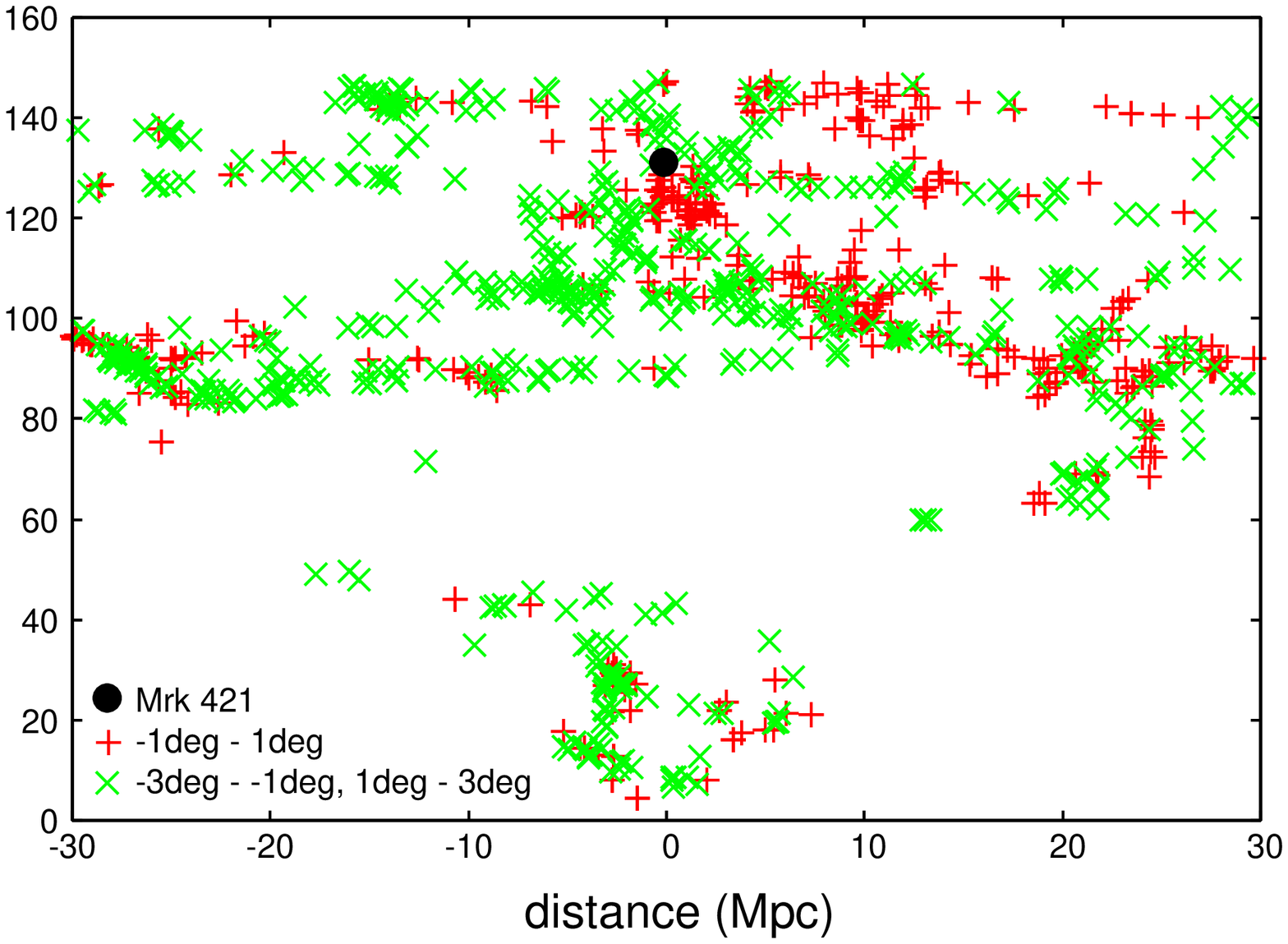}
\caption{Distribution of galaxy locations between our Galaxy (origin) and Mrk 421 (filled circle) mapped by SDSS, projected onto the plane defined by declination 38.2$^{\circ}$ that contains Mrk 421. The symbols represent galaxies within 1 degree (plus) and 1-3 degrees away (cross) from this plane in the sky, respectively. At 100 Mpc distance, 1 degree corresponds to $\sim$1.7 Mpc.}
\label{fig:Mrk421}
\end{figure}

In our previous study \cite{Takahashi3}, we focused on a specific TeV flare of Mrk 501 observed in 2009 by the VERITAS (Very Energetic Radiation Imaging Telescope Array System) and MAGIC (Major Atmospheric Gamma-ray Imaging Cherenkov) telescopes. Comparing the expected light curves of the pair echo from the flare and the concurrent quiescent emission with simultaneous {\it Fermi} observations, we obtained a lower bound on the IGMF amplitude of $B > 10^{-20}~{\rm G}$ at $90\%$ confidence level assuming a field coherence length of 1 kpc. This was obtained with minimal assumptions about the primary TeV emission during unobserved periods or spectral bands, and can be considered more robust in comparison with previous studies \citep{Neronov3,AndoKusenko,Tavecchio1,Tavecchio2,Dolag,Dermer,Neronov2,Taylor,Arlen}.

Here we focus on the TeV blazar Mrk 421 located at $z = 0.031$. As seen in Fig. \ref{fig:Mrk421}, maps of the local galaxy distribution from the Sloan Digital Sky Survey reveal that a large void lies between our galaxy and the supercluster containing Mrk 421 \cite{Abazajian,Blanton}. This is also seen to be the case for Mrk 501. Thus, Mrk 421 is a desirable target for probing IGMFs. Mrk 421 has been monitored continuously at TeV energies by the ARGO-YBJ experiment over the period from 2007 November to 2010 February \citep{Bartoli} (hereafter B11), during which many flares were observed so that more statistically significant bounds on IGMFs can be expected. Note that compared with Cherenkov telescopes, such air shower detectors have a much higher duty cycle and allow uninterrupted long-term observations, albeit at lower sensitivity.

\section{TeV and GeV Emission from Mrk 421}

\begin{figure}[t]
\epsscale{1}
\plotone{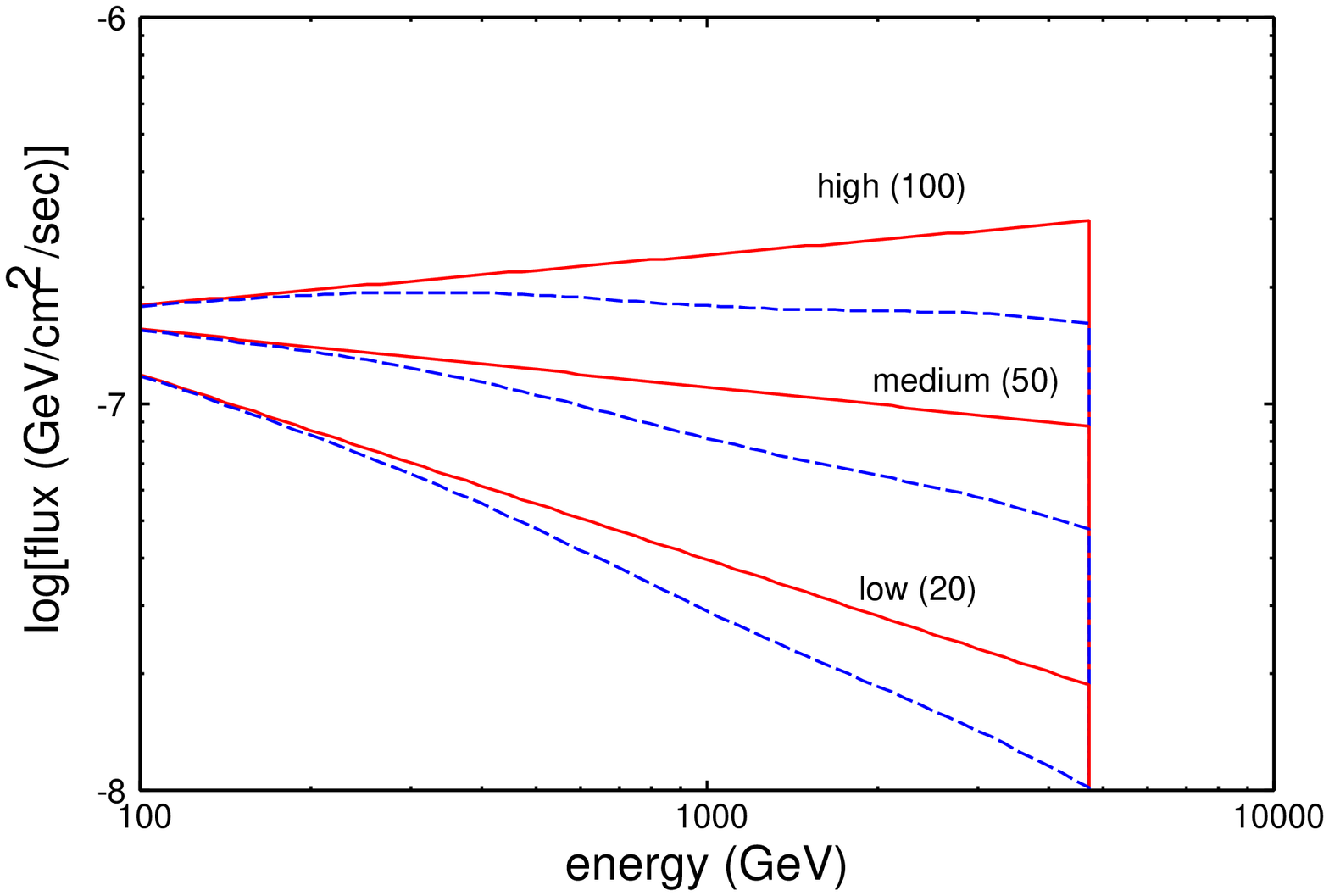}
\caption{Exemplary TeV spectra of Mrk 421 for the high, medium and low states (from top to bottom) with selected values of the the daily TeV counts in parentheses, with (solid) and without (dashed) intergalactic $\gamma\gamma$ absorption.
}
\label{fig:spectrum}
\end{figure}

First we discuss the TeV spectrum and light curve of Mrk 421 with which we evaluate the pair echo. In B11, the daily fluxes at energies above $0.3~{\rm TeV}$ are presented for approximately 850 days. For some days, negative numbers are reported that are presumably caused by systematic errors, and we simply set them to zero. Although the spectra are not available separately for each day, average spectra were derived for four different flux states based on the X-ray count rate. Since the TeV flux was shown to be tightly correlated with that in X-rays, here we choose to define three flux states according to the daily TeV counts, ``high" (count $> 80$), ``medium" ($40 <$ count $< 80$) and ``low" (count $< 40$), which correspond respectively to the X-ray flux levels 4, 3 and 1+2 of B11. Note that levels 1 and 2 can be treated together for our purposes as their TeV spectra are very similar. According to the daily flux, we assume that the TeV spectral index for each day takes the average value of the corresponding flux state. We also impose a maximum spectral cutoff at 5 TeV as the highest energy photons detected by ARGO-YBJ, as well as a minimum cutoff at 0.1 TeV. In Fig. \ref{fig:spectrum}, exemplary spectra for the three states are shown, with and without the effects of intergalactic $\gamma\gamma$ absorption using the cosmic infrared background (CIB) model of Franceschini et al. (2008), also adopted in B11. It turns out that the resulting constraints on the IGMF is largely determined by the high state emission, and the low state is of very little relevance.

For GeV gamma-rays, we utilize the data from {\it Fermi} LAT that has been performing continuous observations of Mrk 421 in the survey mode from MJD 54683. We obtained the data through the {\it Fermi} Science Support Center (FSSC) and adopt the standard analysis tools provided by the FSSC. In our analysis, we divide the energy band in three, that is, 100 MeV-1 GeV, 1-10 GeV, and $>$10 GeV, and derive flux probability distribution functions for each day assuming Poisson statistics. Because the statistics is small for this short interval (1-day bins), we adopt the aperture photometry method where we count events located within 1 degree from the source. Note that we can neglect the background events above 1 GeV for this timescale at the high Galactic latitude of Mrk 421. Below we use data during MJD 54683 - 55255, when both TeV and GeV observations were performed, focusing on the energy range of $1-10~{\rm GeV}$ where {\it Fermi} LAT is most sensitive and the strongest constraints on the pair echo can be obtained.

\section{Pair Echo}

We summarize briefly the basic physics of pair echos (for details, see e.g. Ichiki et al. 2008 and Takahashi et al. 2012). The mean free path of primary gamma-rays with energy $E_{\gamma} \gtrsim 1 ~ {\rm TeV}$ for $\gamma\gamma$ interactions with the CIB is
$
\lambda_{\gamma \gamma}
= 1/(0.26 \sigma_T n_{\rm IR})
= 190~{\rm Mpc}~( n_{\rm IR}/0.01 ~ {\rm cm}^{-3} )^{-1} ,
$
where $\sigma_T$ is the Thomson cross section and $n_{\rm IR}$ is the number density of relevant CIB photons. The interaction results in an $e^-e^+$ pair with energy $E_e \approx E_{\gamma}/2$, which can then inverse-Compton (IC) upscatter ambient CMB photons to produce the pair echo, that is, secondary gamma rays with energy
$
\langle E_{\rm echo} \rangle
= 2.7 T_{\rm CMB} \gamma_e^2
= 2.5~{\rm GeV}~
  ( E_{\gamma}/2 ~ {\rm TeV} )^2,
$
where $T_{\rm CMB} = 2.7~{\rm K}$ is the CMB temperature and $\gamma_e = E_e/m_e c^2$. For primary gamma rays with $E_\gamma \simeq 1 - 5~{\rm TeV}$, $E_{\rm echo} \simeq 1 - 10~{\rm GeV}$. As long as plasma instabilities are unimportant (see below), the pairs continue successive IC scattering until they lose a large fraction of their energy over a length scale
$
\lambda_{\rm IC, cool}
= 3 m_e^2/(4 E_e \sigma_T U_{\rm CMB})
= 350 ~ {\rm kpc}~( E_e/1 ~ {\rm TeV} )^{-1},
$
where $U_{\rm CMB}$ is the CMB energy density. Comparing typical values for $\lambda_{\gamma \gamma}$ and $\lambda_{\rm IC, cool}$, we see that the pairs are generated mostly far away from the source, and then cool over a much smaller scale. Thus, for Mrk 421, the pairs are likely to be produced deep inside and propagate only within the large, intervening void (Fig. \ref{fig:Mrk421}).

It has been suggested recently that rather than IC cooling, the beam of the $\gamma\gamma$-produced pairs may lose much of their energy by heating the intergalactic gas through two-stream-like plasma instabilities \citep{Broderick,Schlickeiser}. If true, it may considerably reduce the pair echo signal, while causing some non-trivial consequences for the evolution of galaxies and the intergalactic medium \citep{Chang,Pfrommer}. However, the actual efficiency and eventual fate of such instabilities has been debated \citep{Miniati} and is highly uncertain at the moment. Below, we proceed on the assumption that such instabilities are insignificant.

A crucial attribute of the pair echo is the time delay compared with the primary gamma-rays, caused by two effects. One is the angular spreading inherent in the pair production and IC scattering processes, for which the typical delay time
$
\Delta t_{\rm ang}
= (\lambda_{\gamma\gamma} + \lambda_{\rm IC, cool})/2 \gamma_e^2
\approx 3 \times 10^3 ~ {\rm sec} ~ (E_{\rm echo}/1 ~ {\rm GeV})^{-1}
        (n_{\rm IR}/0.01 ~ {\rm cm}^{-3})^{-1}
$
(Ichiki et al. 2008). The second is deflections of the pairs in the IGMF with typical delay time
$
\Delta t_{\rm B}
= (\lambda_{\gamma \gamma} + \lambda_{\rm IC, cool})
  \langle \theta_{\rm B}^2 \rangle /2
$,
where
$
\langle \theta_{\rm B}^2 \rangle^{1/2}
= \max [\lambda_{\rm IC, cool}/r_{\rm L}, (\lambda_{\rm IC, cool} r_{\rm coh}/6)^{1/2} / r_{\rm L}]
$
is the typical deflection angle, $r_{\rm L}$ the Larmor radius and $r_{\rm coh}$ the coherence length of the IGMF. If $r_{\rm coh} \ll \lambda_{\rm IC, cool}$, that is, the IGMF is sufficiently tangled on the IC cooling scale,
\begin{eqnarray}
\Delta t_{\rm B}
&\approx& 2 \times 10^4 ~ {\rm sec} ~
(E_{\rm echo}/1 ~ {\rm GeV})^{-3/2}  (B / 10^{-19} ~ {\rm G})^2
\nonumber \\
&& \times (r_{\rm coh}/1 ~ {\rm kpc}) (n_{\rm IR}/0.01 ~ {\rm cm}^{-3})^{-1},
\end{eqnarray}
where $B$ is the field amplitude. Hereafter we take a fiducial value $r_{\rm coh} = 1~{\rm kpc}$ (see e.g. Langer et al. 2005), although the results can be trivially scaled for other values as it always occurs in the combination $B^2 r_{\rm coh}$ if $r_{\rm coh} \lesssim \lambda_{\rm IC, cool}$. The total delay time is approximately $\Delta t = \Delta t_{\rm ang} + \Delta t_{\rm B}$, and the magnetic field properties are reflected in the delay as long as $\Delta t_{\rm ang} \lesssim \Delta t_{\rm B}$.

To calculate the pair echo spectra and light curves, we follow Ichiki et al. (2008). First, the time-integrated flux of secondary pairs is
\begin{equation}
\frac{dN_{e,{\rm 0}}}{d\gamma_e} (\gamma_e)
= 4 m_e
  \frac{dN_{\gamma}}{dE_{\rm \gamma}}(E_{\gamma} = 2 m_e \gamma_e)
  \left[1-e^{-\tau_{\gamma \gamma}(E_{\gamma} = 2 \gamma_e m_e)}\right],
\label{eq:dN0dgamma}
\end{equation}
where $dN_{\gamma}/dE_{\gamma}$ is the primary gamma-ray fluence and $\tau_{\gamma \gamma}(E_\gamma)$ is the $\gamma\gamma$ optical depth in the CIB. The time-dependent pair-echo spectrum is
\begin{equation}
\frac{d^2 N_{\rm echo}}{dt dE_\gamma}
= \int d\gamma_e \frac{dN_e}{d{\gamma_e}}
  \frac{d^2 N_{\rm IC}}{dt dE_\gamma},
\end{equation}
where $d^2 N_{\rm IC}/dt dE_\gamma$ is the IC spectrum from a single electron/positron, and $dN_e/d{\gamma_e}$ is the total flux of pairs relevant for the echo gamma-rays observed at time $t$. This formalism was extended to account for the finite probability of pair production near the observer in Takahashi et al. (2012) (see also Dai et al. 2002). 

\begin{figure}[t]
\epsscale{1}
\plotone{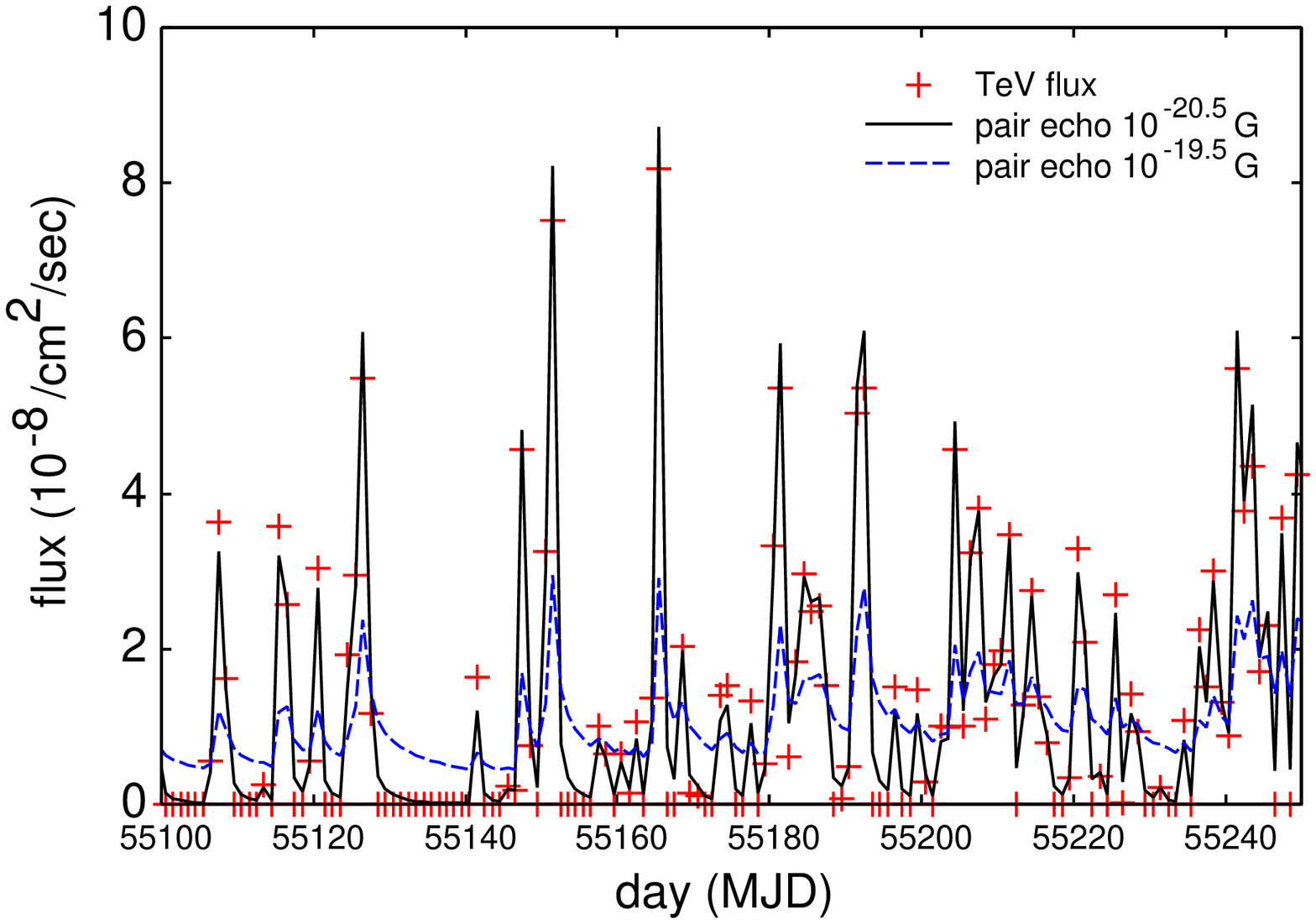}
\caption{Expected light curves of the pair echo at $1-10~{\rm GeV}$ for $B = 10^{-20.5} ~{\rm G}$ (solid) and $B = 10^{-19.5}~{\rm G}$ (dashed) compared with the observed TeV light curve in arbitrary units (crosses).}
\label{fig:lc_TeV-pa2}
\end{figure}

Weaker IGMFs generally lead to higher echo fluxes, as long as the time delay is dominated by $\Delta t_{\rm B}$ rather than $\Delta t_{\rm ang}$. For $r_{\rm coh} = 1~{\rm kpc}$, $\Delta t_{\rm B}$ is of the same order as $\Delta t_{\rm ang}$ if $B \sim 10^{-20}~{\rm G}$. Fig. \ref{fig:lc_TeV-pa2} compares the daily TeV counts with the $1-10~{\rm GeV}$ light curves of the pair echo during a period of 150 days for two values of $B$. While the two are generally correlated, for weaker IGMF, the peak flux of the echo is larger and its response to the primary emission is quicker. Although the magnetic deflection implies that the pair echo emission should also be spatially extended around the primary source, the extension is much smaller than the {\it Fermi} angular resolution and can be neglected for the field strengths of $B \sim 10^{-20}~{\rm G}$ considered here.

\section{Statistical Analysis}

We now compare the expected pair echo with the {\it Fermi}-LAT data and derive constraints on the IGMF. Compared with our previous paper \citep{Takahashi3}, we have a much greater number of independent flux bins (each representing the daily count), so a more sophisticated method of deriving the constraints is necessary. First, we compute the probability $P_i$ that a specific value of the the IGMF amplitude is excluded by the $i$-th flux bin, using the probability distribution function of the true flux obtained from the {\it Fermi}-LAT observation. Then, we combine the probabilities to derive the total probability $P_{\rm tot}$ using meta-analysis.

\begin{figure}[t]
\epsscale{1}
\plotone{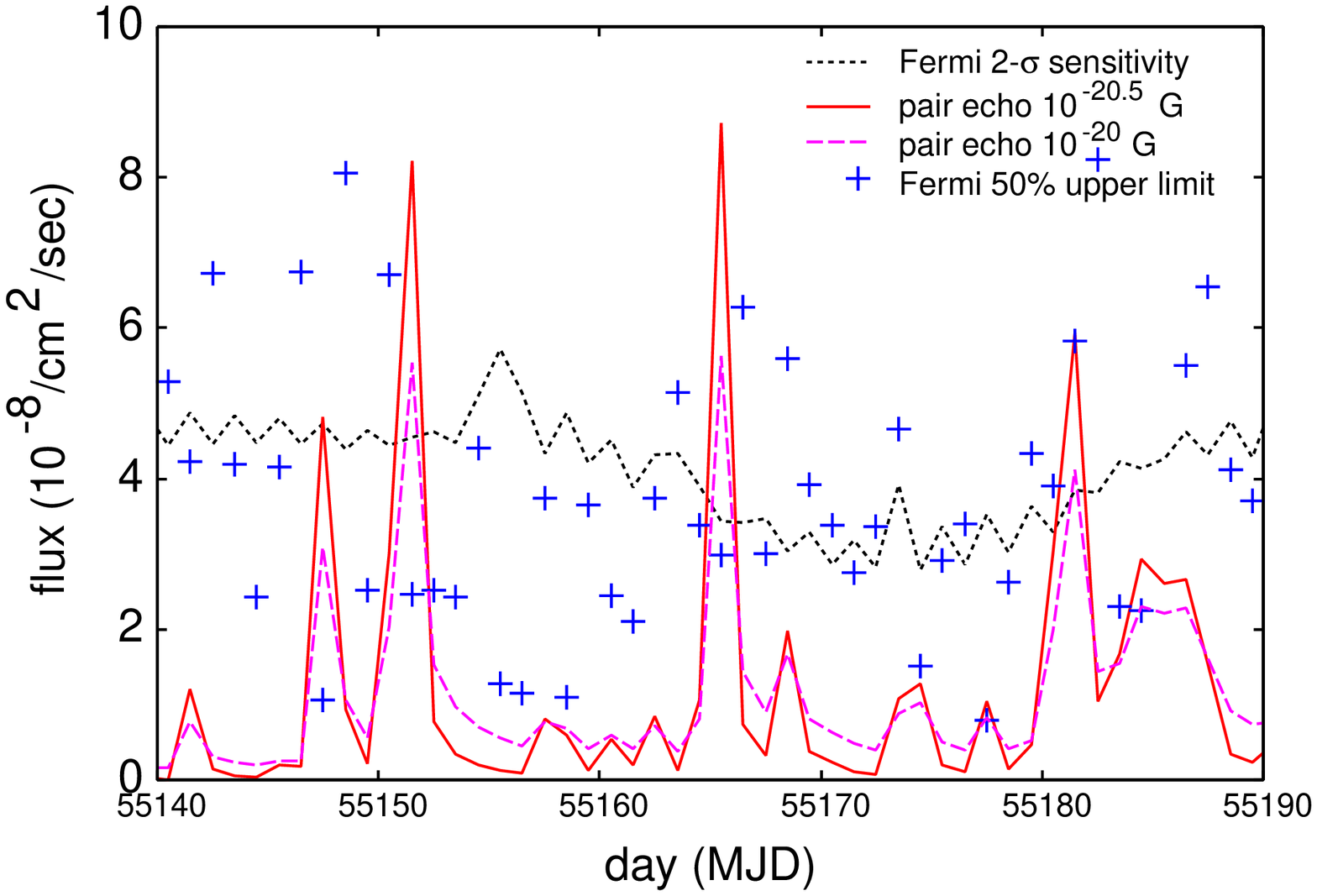}
\caption{Daily {\it Fermi}-LAT 2-$\sigma$ sensitivity (dotted), pair echo light curves for $B = 10^{-20.5}~{\rm G}$ (solid) and $B = 10^{-20}~{\rm G}$ (dashed), and {\it Fermi} $50\%$ confidence upper limits (crosses), all at $1-10~{\rm GeV}$.}
\label{fig:lc_sensitivity-UL}
\end{figure}

Note that it would not be appropriate to simply combine such probabilities for all bins. If the TeV flux for the $i$-th bin is low enough for the expected echo flux to be below the {\it Fermi} sensitivity for that bin, the probability $P_i$ would be small, irrespective of $B$. If we combine all such probabilities, the total probability $P_{\rm tot}$ can become so small that no constraints on $B$ can be obtained, even if some values of $P_i$ are sufficiently large for bins during TeV flares. Thus, we must select data bins for which the expected echo flux would be detectable by {\it Fermi}, depending on the assumed value of $B$. As explained above, larger $B$ results in a weaker echo that can only be detected for bins with higher TeV flux, so the number of such bins will be smaller. Here we set this selection threshold such that the echo flux exceeds the 2-$\sigma$ sensitivity of {\it Fermi}-LAT. In Fig. \ref{fig:lc_sensitivity-UL}, this is compared with the echo light curves for $B = 10^{-20.5}~{\rm G}$ and $10^{-20}~{\rm G}$ at $1-10~{\rm GeV}$ during a particular 50-day period (only a small fraction of the entire data set). Here 4 and 3 bins exceed the {\it Fermi}-LAT sensitivity for $B = 10^{-20.5}~{\rm G}$ and $10^{-20}~{\rm G}$, respectively, which correspond to large TeV flares as seen in Fig. \ref{fig:lc_TeV-pa2}.

Fig. \ref{fig:lc_sensitivity-UL} also plots the $50\%$ confidence {\it Fermi}-LAT upper limits on the daily flux. For the first flare (MJD 55147), the expected pair-echo flux for $B = 10^{-20.5}~{\rm G}$ greatly exceeds the upper limit, and the probability that this value of $B$ is excluded is very large. Although that for $B = 10^{-20}~{\rm G}$ also exceeds the limit, it does not reach the 2-$\sigma$ sensitivity, so the bin is not counted to compute $P_{\rm tot}$ for this $B$ value. For the second (MJD 55152) and third (MJD 55166) flares, the echo fluxes surpass the upper limits as well as the sensitivity for both $B = 10^{-20.5}~{\rm G}$ and $10^{-20}~{\rm G}$. For the fourth flare (MJD 55182), the echo flux for $B = 10^{-20.5}~{\rm G}$ is comparable to the $50\%$ confidence upper limit, neither favoring nor excluding this $B$ value, whereas that for $B = 10^{-20}~{\rm G}$ is not constrained 
by the limit and this $B$ value remains allowed.

We now consider the probability distribution function of the true flux and calculate the probability $P_i$ that it is less than the expected pair-echo flux for the $i$-th bin. To combine $P_i$, we use the inverse normal method, a type of meta-analysis. First, we derive the Z value of the normal distribution for the $i$-th bin, $Z_i$, which is the percentile (point) of the one-sided P value $P_i$. Note that $Z_i$ is negative if $P_i < 0.5$. Next, we compute the total Z value $Z_{\rm tot}$ as
\begin{equation}
Z_{\rm tot} = \frac{1}{\sqrt{N}} \sum_{i=1}^{N} Z_i,
\label{eq:meta}
\end{equation}
where $N$ is the number of the selected bins. Finally, we derive the one-sided P value $P_{\rm tot}$ of the normal distribution that corresponds to the above $Z_{\rm tot}$. We can interpret $P_{\rm tot}$ such that the assumed value of $B$ is excluded at a confidence level of $P_{\rm tot}$.

\begin{figure}[t]
\epsscale{1}
\plotone{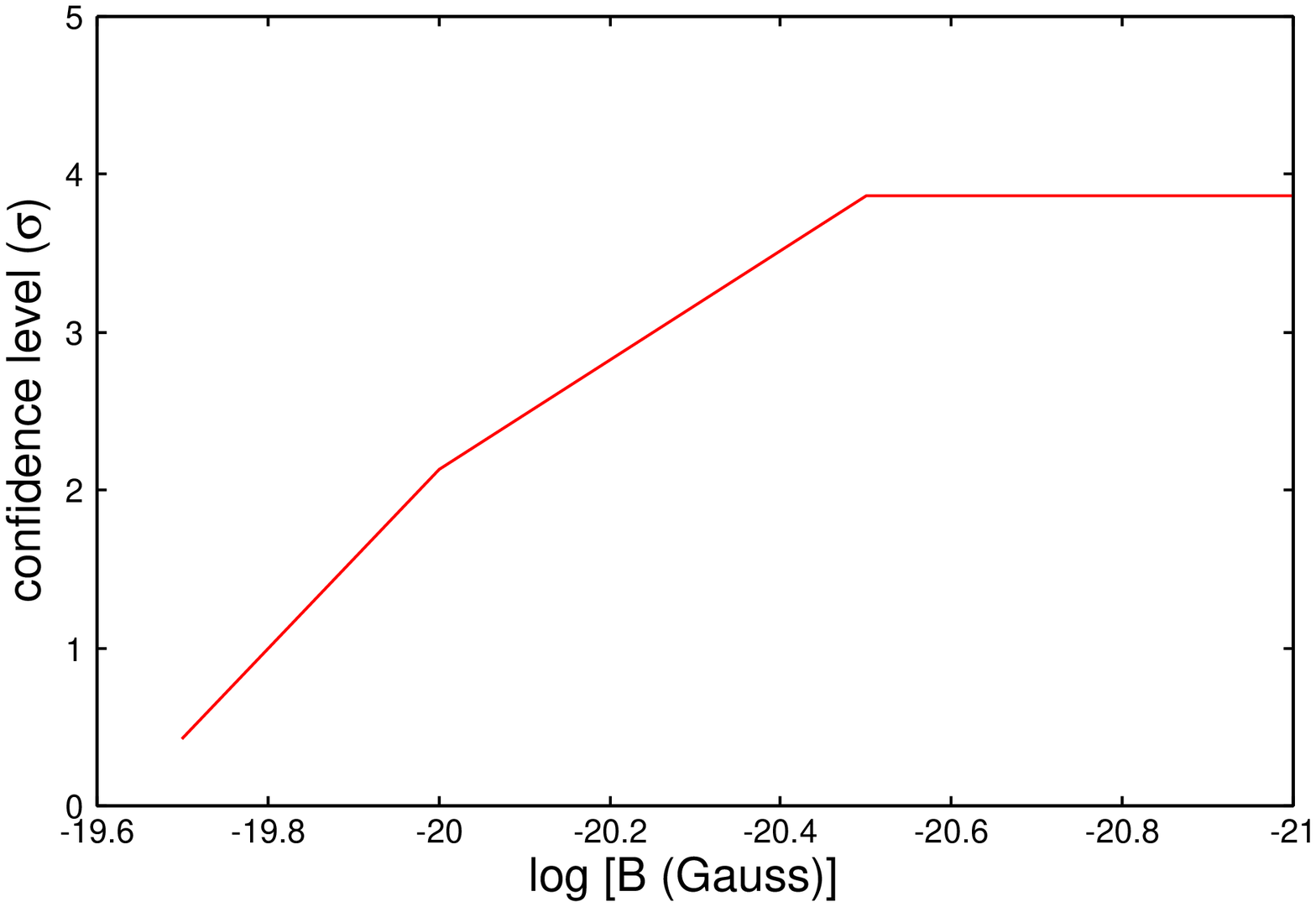}
\caption{The IGMF strength $B$ versus $Z_{\rm tot}$ that describes the confidence level that a value of $B$ is excluded by the {\it Fermi} data.}
\label{fig:constraint}
\end{figure}

Fig. \ref{fig:constraint} shows $Z_{\rm tot}$ as a function of $B$. For $B \leq 10^{-20.5}~{\rm G}$, the delay time of the pair echo is determined by angular spreading and becomes independent of $B$. Such weak IGMFs including $B=0$ is excluded by about 4-$\sigma$ significance. The significance decreases for larger $B$, and no constraints are obtained for $B \gtrsim 10^{-19.7}~{\rm G}$.
This is a consequence of the lack of any time bins for which the pair-echo flux exceeds the 2-$\sigma$ {\it Fermi}-LAT sensitivity when $B \geq 10^{-19.5}~{\rm G}$.

Here we have not considered emission components other than the pair echo in the GeV band. In reality, there is likely to be primary GeV emission from the blazar, and possibly also other types of secondary GeV emission (e.g. Essey et al. 2011). If such components can be reliably accounted for, stronger upper limits on the pair echo and hence stronger lower bounds on the IGMF would be obtainable from the same {\it Fermi} data.

\section{Discussion and Summary}

Using data from long-term, simultaneous GeV-TeV observations of Mrk 421 by {\it Fermi}-LAT and ARGO-YBJ, we have constrained the flux of secondary pair echos and derived lower bounds on the IGMF strength in the large void region lying between our Galaxy and Mrk 421. This was done by: 1) calculating the daily pair-echo flux from the TeV data over 600 days, 2) selecting the dates where the expected pair-echo flux exceeds the {\it Fermi}-LAT 2-$\sigma$ sensitivity, 3) computing the probability that an assumed value of the IGMF is excluded by the {\it Fermi}-LAT data for each date, and 4) combining these probabilities to derive the total probability using the inverse normal method. Consequently, as long as plasma instabilities are inconsequential, IGMFs weaker than $10^{-20.5}~{\rm G}$ are excluded by about 4-$\sigma$ for a field coherence length of 1 kpc. For general values of $r_{\rm coh}$, the derived constraint is $B \gtrsim 10^{-22}~\max[(r_{\rm coh}/350~{\rm kpc})^{-1/2},1]~{\rm G}$, where the latter case corresponds to IGMFs that are coherent over the IC cooling length.

Improving on our previous analysis using Mrk 501 \citep{Takahashi3}, no assumptions are made here concerning the TeV emission during unobserved periods. The obtained constraints are thus more robust than from other studies, particularly those based on limits to the spatially-extended halo emission from secondary pairs that inevitably involves very long time delays, often longer than the typical lifetimes of blazars. Although the value of the lower limit obtained here is similar to our previous work, the statistical significance has increased remarkably, from less than 2-$\sigma$ to about 4-$\sigma$, thanks to the much larger data and improved statistical analysis.

In our study, the errors in the TeV flux, which propagate to the errors in the expected pair-echo flux, have not been considered. However, assuming a Gaussian distribution for the errors, the probability that the true echo flux is larger (or smaller) than the central value is $50\%$, so the errors in the expected echo flux should cancel out among the data bins to some extent and is unlikely to affect the total $P$ value significantly.

Here we have used the {\it Fermi}-LAT data only as daily upper limits to the GeV fluxes. Because the pair-echo flux is strongly dependent on the TeV flux, we can obtain potentially tighter constraints on the IGMF by investigating statistical correlations between the {\it Fermi}-LAT data and the ARGO-YBJ data. This will be presented elsewhere in the near future.

\acknowledgments

We thank Songzhan Chen who kindly provided us with the ARGO-YBJ data. This work is supported in part by the Grant-in-Aid from the Ministry of Education, Culture, Sports, Science and Technology (MEXT) of Japan, No. 23740179, No. 24111710 and No. 24340048 (KT), No. 22540315 (MM), No. 24340048 (KI) and No. 22540278 (SI), and by the Grant-in-Aid for the global COE program ``Quest for Fundamental Principles in the Universe: from Particles to the Solar System and the Cosmos'' at Nagoya University from MEXT of Japan. The work of HT is supported by JSPS fellowship.

\end{document}